\documentclass[numcites
  ] {aipproc}

\layoutstyle{6x9}

\setcitestyle{numbers}

\begin{document}

\title{Hard Probe of Soft Matter Geometry and Fluctuations from RHIC to LHC}


\author{Jinfeng Liao}{
  address={Physics Department and Center for Exploration of Energy and Matter,
Indiana University, \\ 2401 N Milo B. Sampson Lane, Bloomington, IN 47408, USA.},
altaddress={RIKEN BNL Research Center, Bldg. 510A, Brookhaven National Laboratory, \\ Upton, NY 11973, USA.}
}

\begin{abstract}
 We report results on event-by-event hard probe of soft matter geometry and 
fluctuations in heavy ion collisions. Geometric data ($v_2$ of high $p_t$ hadrons) 
from RHIC plus LHC clearly favors jet ``monography'' model with strong near-Tc enhancement of jet-medium interaction strength which also implies a less opaque medium at LHC.  We also quantify the jet responses to all harmonic anisotropy  $v_n$($n=1,2,3,4,5,6$)  and their manifestation in hard-soft azimuthal correlations.   
\end{abstract}

\maketitle


High energy heavy ion collisions have provided the unique laboratory access to the hot deconfined QCD matter. 
With extensive measurements over the past decade at RHIC and immense body of LHC data freshly arriving, we are at an age of quantitatively studying the ``condensed matter physics of QCD'' for advancing our knowledge of how QCD operates in Nature. A powerful ``imaging'' tool for the created matter is the hard probe produced in the initial binary hard collisions. Such a partonic jet  experiences multiple collisions in the dense medium and loses energy (i.e. jet quenching), leading to measurable differences in the final hadrons as compared with e.g. in $pp$ collisions and providing a way of learning about medium properties  and jet-medium interactions. 

This study focuses on the geometric aspects of jet quenching. The hot medium created in a heavy ion collision event is generally anisotropic in the transverse plane, therefore high energy partons traversing the medium along different azimuthal directions will ``see'' different medium thickness and thus lose different amount of energy.  That will lead to a measurable anisotropy in the nuclear modification factor $R_{AA}(\phi)$. The so-obtained $R_{AA}(\phi)$ may be further Fourier decomposed  as: $R_{AA}(\phi)=R_{AA}\left(1+2\sum_{n=1}  v_{n}\cos[n(\phi-\Psi^J_{n})]\right)$. 
The overall quenching $R_{AA}$ as well as the azimuthal harmonics $v_n$ (for high $p_t$ hadrons) and the corresponding n-axis $\Psi^J_{n}$ can then be determined from the above decomposition. The second harmonic $v_2$ is the robustest and reflects the hard probe of anisotropy both from {\em geometry and fluctuation}, which is very sensitive to the underlying dynamics of jet energy loss, in particular its dependence on the path-length and medium-density \citep{Gyulassy:2000gk,Shuryak:2001me,Drees:2003zh,Liao:2008dk,Jia:2010ee,Jia:2011pi,Betz:2011tu,Liao:2011kr,Rodriguez:2010di} . More recent studies \citep{Zhang:2012mi} have extended to quantify  the various other harmonics   which arise from jet responses to strong event-by-event {\em fluctuations}  and provide further insights into the initial conditions. 

\begin{figure}
		\includegraphics[width=5.5cm]{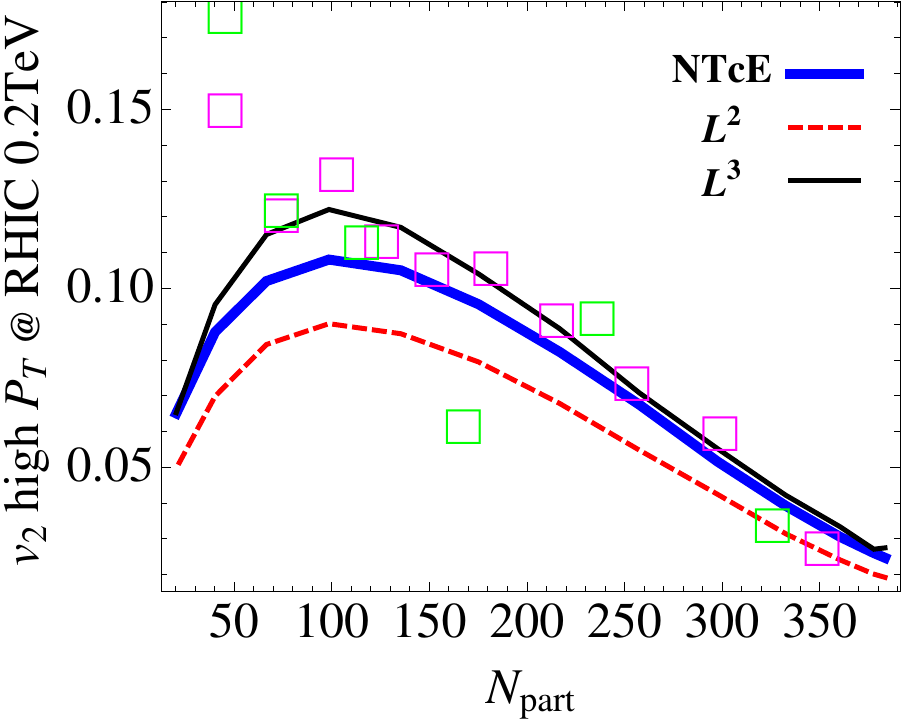} \hspace{0.5cm}
		\includegraphics[width=5.5cm]{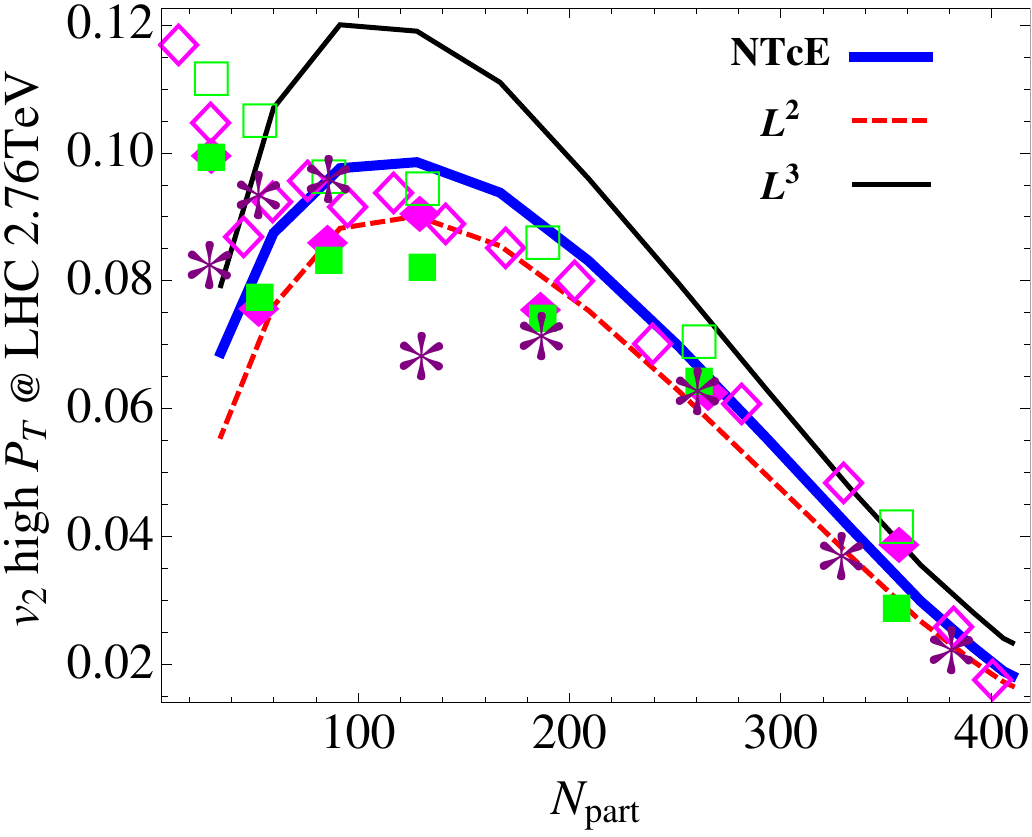}
		\caption{(color online) The $v_2$ of high $p_t$ hadrons at RHIC $0.2\rm\, TeV$(left) and LHC $2.76\rm\, TeV$(right) collisions computed from NTcE (thick solid blue),  $\mathrm{L}^{2}$ (dashed red) as well as $\mathrm{L}^{3}$ (thin solid black) models. The results are compared with various data from RHIC plus LHC (see \citep{RHIC,LHC,Zhang_Liao} for details).}
		\label{V2_comparison}
\end{figure}

Studies of $v_2$ for high $p_t$ hadrons at RHIC, however, revealed a clear discrepancy between various model results and  data  \citep{Gyulassy:2000gk,Shuryak:2001me,RHIC}.  Efforts toward reconciling $R_{AA}$ and $v_2$ at high $p_t$ succeeded only with a more radical proposal in \citep{Liao:2008dk}  that the jet-medium interaction is strongly enhanced in the near-$T_c$ plasma. Such near-$T_c$ enhancement of jet quenching, in analogy with ``critical opalescence'', is motivated by the scenario  that the near-$T_c$ plasma is an emergent matter dominated by dense and light (chromo-)magnetic monopoles\citep{Liao:2006ry}. Lately an AdS/CFT-motivated model with cubic path-length dependence also described $v_2$ at  high $p_t$ for RHIC \citep{Jia:2010ee}, presumably because it effectively enhances the late time quenching that mimics the near-$T_c$ enhancement. 

Now that with available geometric data from RHIC plus LHC, it is tempting to further discriminate models and pin down the  path-length and matter-density dependences of jet quenching. To do that we here use  and compare three classes of geometric models (see \citep{Liao:2011kr,Zhang:2012mi,Zhang_Liao} for details): the NTcE model with strong near-$T_c$ enhancement of jet-medium interaction strength $\kappa(s)$ ($s$ the entropy density) and quadratic path-length dependence, in contrast with the $\mathrm{L}^{2}$ model and the $\mathrm{L}^{3}$ model that both have a constant jet-medium interaction strength $\kappa_0$ while have quadratic and cubic path-length dependence, respectively. (Note the strength $\kappa$ is without the trivial density factor and roughly corresponds to $\hat{q}/s$ in certain models.) We've extracted $R_{AA}(\phi)$ and the $v_n$ coefficients event-by-event with a Monte-Carlo Glauber simulation for all three models \citep{Zhang_Liao}. The results for $v_2$ are shown in Fig.\ref{V2_comparison} and compared with data: evidently the $L^2$ model (``tomography'') fails at RHIC and the $L^3$ model (``holography'') fails at LHC, while only the NTcE model (``monography'') is consistent with all data.

\begin{figure}
		\includegraphics[width=5.cm]{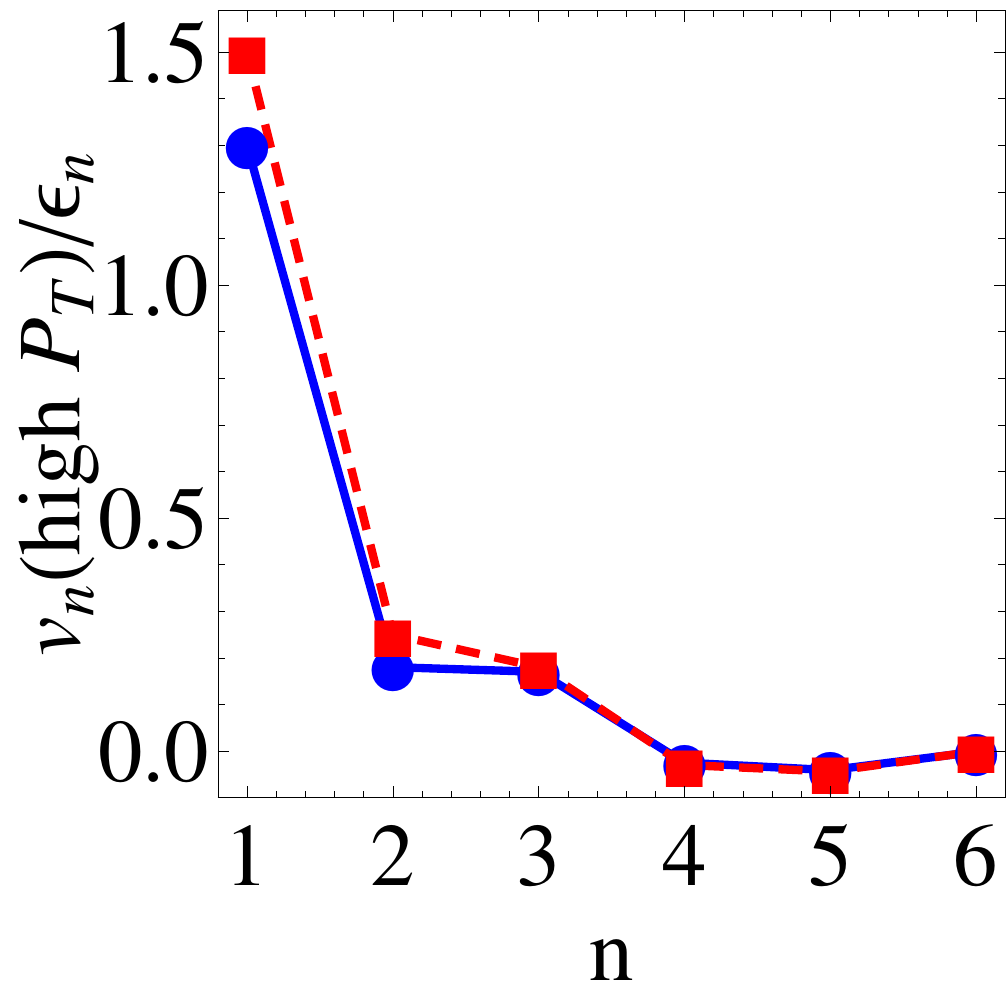} \hspace{0.5cm}
		\includegraphics[width=5.4cm]{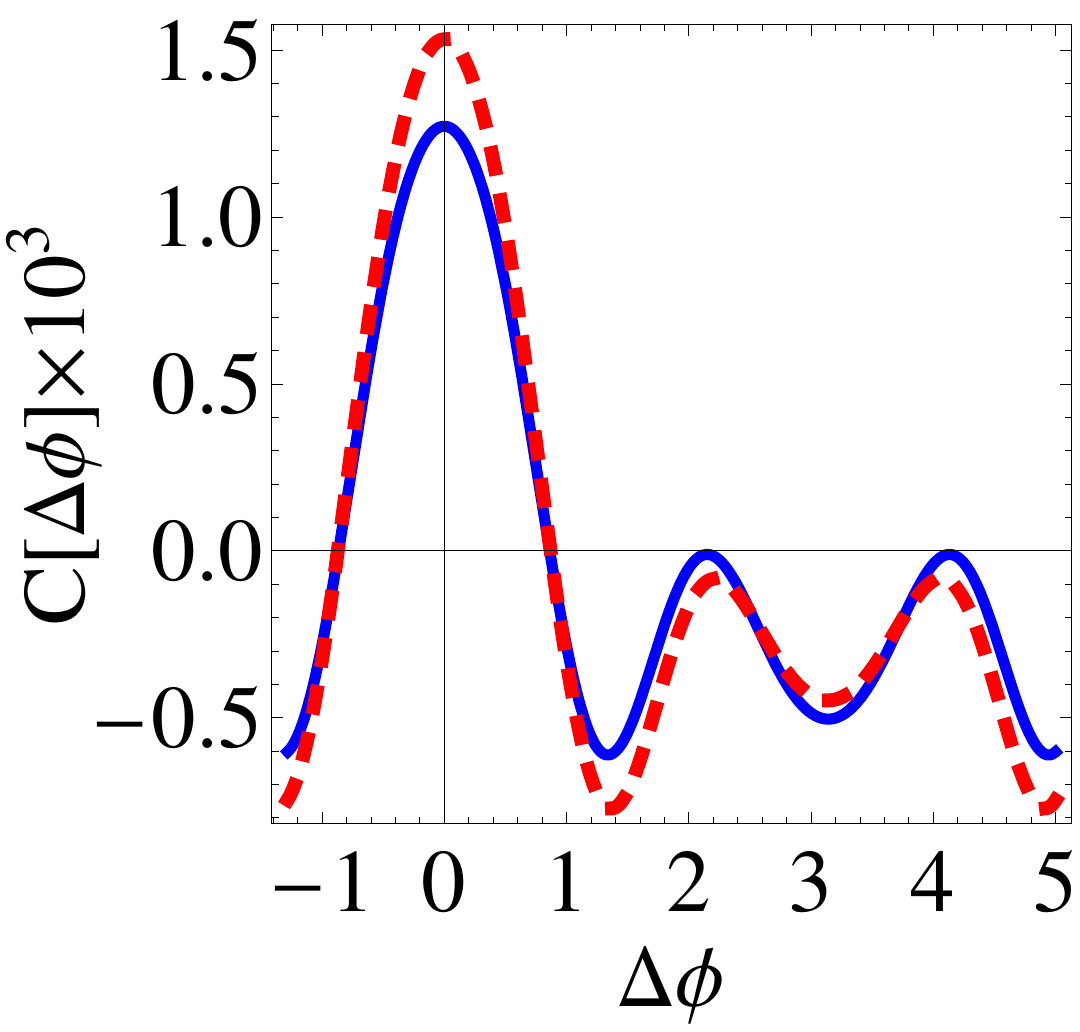}
		\caption{(color online) (left) Jet response to harmonic fluctuations $v_n/\epsilon_n$ for $n=1$ to $6$ and (right) the hard-soft azimuthal correlation arising from the fluctuations in initial condition (see \citep{Zhang:2012mi} for details).	}	\label{Vn}
\end{figure}
 
Hard probe of fluctuations is  a quite new and interesting direction which may lead toward independent and complimentary constraints on the initial condition, in addition to that from bulk collective expansion. In \citep{Zhang:2012mi} we've first attempted  a systematical quantification of the jet response to harmonic fluctuations up to $n=6$: see Fig.\ref{Vn}(left) for $v_n/\epsilon_n$.      Such results further allow us to show how the hard and soft sectors get mutually correlated on an event-by-event basis through their common correlations to the same initial condition with fluctuating geometry. Schematically the correlation can be expressed as $C[\Delta \phi] \sim \sum_{n=1,2,3,...} 2\, < v^{h}_n v^{s}_n>\, \cos(n\Delta\phi)$ and  is demonstrated in Fig.\ref{Vn}(right) based on our computed jet response $v^{h}_n$ and the hydro response $v^{s}_n$ from literature: interesting features like the near-side ``hard ridge'' and away-side ``shoulder'' clearly arise.

We end by pointing out that in the ``monography'' picture with strongly peaked jet-medium interaction strength near $T_c$, one naturally expects (on average) a less color-opaque medium created at LHC as compared with that at RHIC \citep{Liao:2008dk,Liao:2011kr}.  A ``less opaque medium'' may be phrased in varied ways, e.g. over-quenching at LHC if scaling up with density the same parameter sets from RHIC, a noticeable reduction of coupling constant directly extracted from LHC as compared with that obtained with same procedure at RHIC, extracted $\hat{q}$ values not scaling with the density increase, etc. A first indication is that even with multiplicity doubled, the central $R_{AA}$ at LHC is close to that at RHIC. Recently multiple indenpendent jet quenching studies (e.g. GLV/WHDG/CUJET, ``polytrope'' models, geometric models, data scaling analysis, even lattice attempts) \citep{Less_Opaque} 
indeed appear to report the coherent message of a less opaque medium at LHC.





\end{document}